\documentclass[conference]{IEEEtran}
\IEEEoverridecommandlockouts
\usepackage{cite}
\usepackage{amsmath,amssymb,amsfonts}
\usepackage{algorithmic}
\usepackage{graphicx}
\usepackage{textcomp}
\usepackage{xcolor}
\usepackage{color}

\def\BibTeX{{\rm B\kern-.05em{\sc i\kern-.025em b}\kern-.08em
    T\kern-.1667em\lower.7ex\hbox{E}\kern-.125emX}}
    
\IEEEoverridecommandlockouts
\IEEEpubid{\begin{minipage}{\textwidth}\ \\[80pt]
        \centering\normalsize{	© 2021 IEEE. Personal use of this material is permitted. Permission from IEEE must be
obtained for all other uses, in any current or future media, including
reprinting/republishing this material for advertising or promotional purposes, creating new
collective works, for resale or redistribution to servers or lists, or reuse of any copyrighted
component of this work in other works.}
\end{minipage}} 
\begin{document}

\title{Similarity Measures for Location-Dependent MMIMO, 5G Base Stations On/Off Switching Using Radio Environment Map  \\

\thanks{The simulations were based on the QCM simulator from
Huawei Technologies Sweden Research Center.
The presented work was funded by the Polish Ministry of Science and Higher Education subvention within the task "New methods of increasing energy and spectral efficiency and localization awareness for mobile systems" in 2020.}
}

\author{\IEEEauthorblockN{ Marcin Hoffmann}
\IEEEauthorblockA{\textit{Institute of Radiocommunications} \\
\textit{Poznań University of Technology}\\
Poznań, Poland \\
marcin.ro.hoffmann@doctorate.put.poznan.pl}
\and
\IEEEauthorblockN{ Paweł Kryszkiewicz}
\IEEEauthorblockA{\textit{Institute of Radiocommunications} \\
\textit{Poznań University of Technology}\\
Poznań, Poland \\
pawel.kryszkiewicz@put.poznan.pl}
}

\maketitle

\begin{abstract}
The Massive Multiple-Input Multiple-Output (MMIMO) technique together with Heterogeneous Network (Het-Net) deployment enables high throughput of 5G and beyond networks. However, a high number of antennas and a high number of Base Stations (BSs) can result in significant power consumption. Previous studies have shown that the energy efficiency (EE) of such a network can be effectively increased by turning off some BSs depending on User Equipments (UEs) positions. Such mapping is obtained by using Reinforcement Learning. Its results are stored in a so-called Radio Environment Map (REM).
However, in a real network, the number of UEs' positions patterns would go to infinity. This paper aims to determine how to match the current set of UEs' positions to the most similar pattern, i.e., providing the same optimal active BSs set, saved in REM. We compare several state-of-the-art distance metrics using a computer simulator: an accurate 3D-Ray-Tracing model of the radio channel and an advanced system-level simulator of MMIMO Het-Net. The results have shown that the so-called Sum of Minimums Distance provides the best matching between REM data and UEs' positions, enabling up to 56\% EE improvement over the scenario without EE optimization.
\end{abstract}


\section{Introduction}

Key enablers for achieving high network throughput in 5G and beyond, are Massive Multiple-Input Multiple-Output (MMIMO), and Heterogeneous Network (Het-Net) design~\cite{dogra2020}. MMIMO provides throughput gain by utilizing large antenna arrays transmitting energy directly to the user equipment (UE), i.e., beamforming. Het-Net stands for the idea where an additional tier of small (e.g., pico) Base Stations (BSs) is densely deployed close to the UEs. It results in macro BS traffic being offloaded and increases network capacity. Although the deployment of both Het-Net and MMIMO can effectively reduce the transmit power in the network, it is not always enough to compensate energy consumption related to additional hardware~\cite{bjornson2015}. Thus, the improvement of 5G and future 6G networks' energy efficiency (EE) is of high importance~\cite{lee2020}. Research has shown that many of the currently deployed BSs are underutilized over long periods of time~\cite{jin2012}. The problem would arise even more while considering the dense deployment of pico BSs. Some of those underutilized BSs can be temporally switched off without affecting UEs' Quality of Service (QoS) in order to improve network EE. There are a number of algorithms relying on standard optimization methods proposed in the literature to solve this problem~\cite{han2016}. However, in most cases, they suffer from simplistic modeling of a wireless network. Proper modeling is even more difficult when MMIMO is used. It is because a MMIMO network consists of several complex functional blocks including precoding, user scheduling, channel estimation, and user-to-BS assignment. As a result, the behavior of the system is hard to predict efficiently. Thus, it is more beneficial to utilize one of the machine learning approaches, e.g., Reinforcement Learning (RL). In RL, a so-called agent interacts with the network and observes the outcome, e.g., EE related to a given set of active BSs. Some algorithms utilizing RL to provide EE gains through BSs on/off switching have already been proposed~\cite{sharma2019, islam2019}. These RL algorithms switch BSs on/off based on bitrates reported by UEs, i.e., the state is defined as a set of UEs bitrates. Because the set of active BSs affects future UEs bitrates this implies a complex sequence of states. 
    
    However, 5G networks are expected to come with accurate localization techniques~\cite{hoffmann2020rtk}. In our previous papers, we have presented the idea of utilizing RL for BSs on/off switching based on UEs' positions to improve network EE~\cite{hoffmann2020, HOFFMANN2021}. We proposed to create a Radio Environment Map (REM), which provides the network with information about an optimal, in terms of EE, set of active BSs related to the current set of UEs' positions. This idea evolved from the earlier works, where REM was used, e.g., to enable opportunistic transmission in licensed frequency bands~\cite{kliks2017}. Originally an entry in REM is created as a tuple of a localization tag, and measured radio link parameter, e.g., Received Signal Strength (RSS). However, the idea can easily be extended to fill REM with any kind of location-dependent data that can be beneficial in terms of network performance~\cite{tengkvist2017}. The proposed mapping between the set of UEs' positions, and active BSs is beneficial in terms of RL application.  
    Because an active BSs set does not affect UEs motion, EE can be optimized independently for every set of UEs' positions. Thus, the full RL problem is simplified to the so-called Contextual Bandit problem. The main focus of \cite{hoffmann2020, HOFFMANN2021} was on the learning procedure and improvement of its speed, i.e., how to design learning so that a minimal number of iterations is required to achieve optimal EE. Thus the algorithm started from an empty REM and the same UEs motion pattern was considered all the time. 
    In a real network, the number of different UEs' positions patterns would go to infinity. As such REM has to be equipped with a similarity measure that allows merging different patterns, reducing the REM size but also improving the convergence speed. 
    
    While \cite{hoffmann2020, HOFFMANN2021} focused mainly on learning from a time perspective, this paper proposes learning optimization from a spatial perspective. 
    Here it is assumed that REM contains some location-dependent knowledge from previous users' patterns and learning. The data in REM is tagged with a set of UEs' locations, and the research issue addressed in this paper is if a new set of UEs positions could be matched to one of the already existing REM entries, and which one should it be. This procedure aims to improve network EE, without the necessity to perform a long learning phase. Several state-of-the-art distance metrics are compared in terms of computer simulations, in order to select the one which efficiently utilizes gains from historical knowledge stored in REM. An advanced system-level simulator of a MMIMO network and a realistic 3D-Ray-Tracing radio channel model are used for this purpose. Moreover, distance metrics are evaluated with two localization methods: almost perfect Real Time Kinematics (RTK), and much less accurate standard Global Positioning System (GPS).                
 
    The paper is organized as follows: The system model is presented in~Sec.~\ref{sec:system_model}. REM and related RL procedures are described in~Sec.~\ref{sec:rem_deployment}. The considered distance metrics are presented in Sec.~\ref{sec:distance_metrics}. The results of computer simulations are discussed in Sec.~\ref{sec:results}. The paper is concluded in~Sec.~\ref{sec:conclusions}. 

\section{System Model} \label{sec:system_model}

In this section, a brief description of the considered system, power modeling, and EE definition is introduced. For more details, we refer the reader to our previous paper~\cite{HOFFMANN2021}.

\subsection{Network Architecture} \label{subsec:network_architecture}

In this paper, we consider a MMIMO Het-Net consisting of one Macro BS (MBS), and $N_{\mathrm{BS}}-1$ Pico BSs (PBSs). The Het-Net is considered to be deployed under urban conditions. MBS and PBSs share a common frequency band and utilize Orthogonal Frequency Division Multiple Access (OFDMA). All BSs support MMIMO, and are equipped with $m \times n$ (columns $\times$ rows) rectangular antenna arrays, exploiting up to above a hundred antenna elements. There are $N_{\mathrm{UE}}$ UEs randomly distributed over the network, moving with speed $v$. The position of the $i$-th UE is reported by standard cellular localization techniques utilizing either RTK or a standard GPS system, as described in 3GPP TS.38.305. The coordinates of the $i$-th UE are given in Cartesian coordinates as $\mathbf{x}_i = [x_i \text{ } y_i]$.  We assume that UE can be served by the network when its RSS is above a given threshold $P_{th}$ for at least one of the BSs.  The MBS is expected to manage the process of location-dependent PBSs switching on/off and is not considered for being switched off. The process of PBSs switching on/off is performed with the use of positions reported by UEs and REM data. To enable easy exchange of information, REM is expected to be deployed on the MBS. The general aim of REM-based PBSs on/off switching procedure is to improve network EE. 

\subsection{Power Model}

To assess network EE, it is crucial to know how much power this network consumes. When considering the MMIMO network, a proper power consumption model must be chosen. It is because of the significant contribution of hardware to the total amount of energy consumption in MMIMO BSs. Such a proper model was found in~\cite{massivemimobook}. We decided to use three major power consumption model components \cite{hoffmann2020, HOFFMANN2021}:
\begin{itemize}
    \item \textbf{Effective Transmitted Power (ETP) $P_{\mathrm{ETP},b}$} being the total power $P_{\mathrm{tx},b}$ radiated by the BS~$b$, and affected by amplifier efficiency~$\eta$: 
    \begin{equation}
        P_{\mathrm{ETP},b} = \frac{P_{\mathrm{tx},b}}{\eta}.
    \end{equation}
    \item \textbf{Transceiver Chains Power (TCP)} $P_{\mathrm{TC},b}$, being the power consumed by the local oscillator $P_{\mathrm{LO}}$, and hardware $P_{\hat{\mathrm{TC}}}$ related to each of BS $b$ antennas $M_b$:
    \begin{equation}
        P_{\mathrm{TC},b} = M_b \cdot P_{\hat{\mathrm{TC}}} + P_{\mathrm{LO}}.
    \end{equation}
    \item \textbf{Fix Power} $P_{\mathrm{fix}}$, is the constant amount of power necessary for, e.g., back-haul signaling and baseband signal processing.
\end{itemize}
By switching BSs off we mean putting them into the stand-by mode in fact. In standby mode, most of the BS hardware components are switched off in order consume only the minimal amount of power denoted as $P_{\mathrm{off}}$. BS in the standby mode can be activated almost immediately, e.g., within $30$~$\mu$s~\cite{frenger2011}. The total amount of power consumed by the system is given as:
\begin{equation} \label{eq:power_consumption}
    P_{tot} = \sum_{b=1}^{N_{\mathrm{BS}}}{P_{tot,b}},
\end{equation}
where
\begin{equation}
    P_{tot,b} =\begin{cases}
    P_{\mathrm{ETP},b} + P_{\mathrm{TC},b} + P_{\mathrm{fix}}, & \text{for active BS,} \\
    P_{\mathrm{off}}, & \text{for BS in stand-by mode}.
    \end{cases}
\end{equation}

\subsection{Energy Efficiency Definitions} \label{subsec:energy_efficiency}

Among several definitions of EE proposed in the literature, the most common is that EE is the ratio between the average UE bitrate and the average power consumed by the network~\cite{tombaz2014}. We propose a similar EE definition, yet using a median of UE bitrate $c_{\mathrm{50}}$ instead of average:
\begin{equation} \label{eq:energy_efficiency}
    EE = \frac{c_{\mathrm{50}}}{P_{\mathrm{avg}}},
\end{equation}
where $P_{\mathrm{avg}}$ is the average power consumption computed according to~\eqref{eq:power_consumption}. This definition of EE provides more fairness and protects UEs with poor radio conditions. The procedure of REM-based BSs on/off switching aims at the maximization of EE given as above over different sets of UEs' positions. However, there is also an additional QoS constraint. Switching off BS cannot cause disconnecting UEs from the network. 

\section{Radio Environment Map \& Reinforce Learning for MMIMO network EE optimization} \label{sec:rem_deployment}
\begin{figure}[htbp]
\centerline{\includegraphics[scale=0.34]{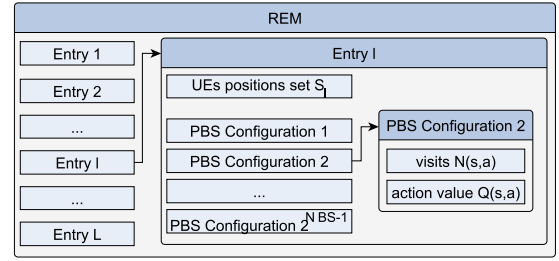}}
\caption{Data structure in REM. While MBS is not considered for switching off, there are $2^{N_{\mathrm{BS}}-1}$ possible active BSs sets.}
\label{fig:rem_data}
\end{figure}

The data stored in REM is organized in entries, and the procedure of EE optimization is performed independently within each entry. 
A REM entry is tagged with a set of UEs positions, i.e., $\mathcal{S}_l:\{ \mathbf{x}_i \}_{i=1}^{N_{\mathrm{UE}}}$ for the $l$-th REM entry. Instead of a single measured value, as in most state-of-the-art REM implementations, our REM entry contains information about network performance for each of $2^{N_{\mathrm{BS}}-1}$ possible on/off configurations of PBSs. The resultant data structure of REM is depicted in Fig.~\ref{fig:rem_data}. The presented REM data structure is designed to support RL, thus related nomenclature is used in Fig.~\ref{fig:rem_data}, e.g., $Q(\mathcal{S}_l,\mathbf{a})$ being the so-called action value. 

\subsection{Reinforcement Learning} \label{subsec:RL}
The general idea of RL is that a so-called agent obtains knowledge about the environment through interaction in discrete time intervals, and observation of outcome, i.e., reward~\cite{Sutton1998}. At the beginning of each interval, the RL agent recognizes the environment state, which is a set of current UEs positions $\mathcal{S}_l$. Later on, the agent interacts with the environment, i.e., it makes an action denoted as $\mathbf{a}$. In our case, action $\mathbf{a}$ represents a set of active PBSs. 
There are several algorithms of obtaining action $\mathbf{a}$, usually in order to balance exploration and exploitation, e.g., $\epsilon$-greedy, Upper Confidence Bound. Some of these algorithms utilize knowledge about the number of times a particular action was chosen in the past $N(\mathcal{S}_l,\mathbf{a})$, which is stored in REM. At the end of the time interval, the environment responds with a reward which is related to EE definition~\eqref{eq:energy_efficiency}:
\begin{equation} \label{eq:reward}
    r(\mathcal{S}_l,\mathbf{a}) = \begin{cases} EE(\mathcal{S}_l,\mathbf{a}), & \text{if } N_{\mathrm{UE}}(\mathcal{S}_l,\mathbf{a}) = N_{\mathrm{UE}}(\mathcal{S}_l,\mathbf{1}), \\
    0, & \text{otherwise,}
    \end{cases}
\end{equation}
where $N_{\mathrm{UE}}(\mathcal{S}_l,\mathbf{a})$ denotes number of UEs connected to network under active PBS set $\mathbf{a}$, and $N_{\mathrm{UE}}(\mathcal{S}_l,\mathbf{1})$ denotes number of UEs that would be connected to network if all PBSs were active, i.e., under action $\mathbf{1}$. Finally, reward value is utilized to update so-called action values $Q(\mathcal{S}_l,\mathbf{a})$ that are stored in REM. Action values are the measure of profitability of choosing particular action being in state $\mathcal{S}_l$, in order to maximize expected reward, i.e, EE. There are also several update rules of action values available, e.g., Q-learning~\cite{Sutton1998}.

\subsection{Utilization of REM data by RL} \label{subsec:utilization_of_rem}

The procedure of REM construction with the use of RL is presented in detail in \cite{hoffmann2020, HOFFMANN2021}. In this paper, we assume that we have REM already filled with $L$ entries, i.e., for L different UEs location patterns the learning was already performed. The current set of UEs positions is denoted as $\mathcal{S}_{\tilde{l}}$. Our target is to specify such a distance metric that finds a REM entry characterized by a set of UEs positions $\mathcal{S}_l$ most similar (in the sense of the optimal set of active PBSs)  to $\mathcal{S}_{\tilde{l}}$:
\begin{equation} \label{eq:state_matching}
    \hat{l} = \arg \min_l{d_f(\mathcal{S}_l,\mathcal{S}_{\tilde{l}})},
\end{equation}
where $d_f(\mathcal{S}_l,\mathcal{S}_{\tilde{l}})$ denotes the distance between $\mathcal{S}_l$ and $\mathcal{S}_{\tilde{l}}$ computed according to metric $f$.
After obtaining the most similar REM entry $\hat{l}$, first, a procedure of Action Space Reduction (ASR) is launched, aiming at the rejection of all PBSs configurations that result in serving fewer UEs, than a configuration with all PBSs turned on~\cite{HOFFMANN2021}. The remaining PBSs configurations formulate action space $\tilde{\mathcal{A}}$. Finally, a greedy action is selected:
\begin{equation} \label{eq:action_selection}
    \mathbf{a}_t = \arg\max_{\mathbf{a} \in \tilde{\mathcal{A}}} Q(\mathcal{S}_{\hat{l}},\mathbf{a})
\end{equation}

\section{Distance Metrics} \label{sec:distance_metrics}

It is expected that the result of~\eqref{eq:state_matching} highly depends on the chosen distance metric $d_f(\cdot,\cdot)$. During our previous studies, we have arbitrarily chosen the Hausdorff Distance for this purpose. However, the Hausdorff Distance is reported as very sensitive to changes in a single UE position, even when the overall structure in the set of UEs' positions remains very similar~\cite{eiter99}. Fortunately, there are several other metrics proposed in the literature~\cite{sherif2017}. 
We decided to compare four distance metrics in total,
being in our opinion the most intuitive and suitable for our problem, e.g., enabling the comparison of two position sets with a different number of UEs. 

\subsection{Hausdorff Distance}
The idea of the Hausdorff Distance is that first, for each UE position $\mathbf{x}_n$ in set $\mathcal{S}_l$, a Euclidean Distance to the closest UE position $\mathbf{x}_m$ in set $\mathcal{S}_{\tilde{l}}$ is computed. Then the maximum value of these distances is taken. Next, the procedure is repeated in reversed order, and the max of the two resultant values is taken. Mathematically, it can be defined as:
\begin{equation} \label{eq:hausdorf_distance}
    d_{hd}(\mathcal{S}_l, \mathcal{S}_{\tilde{l}}) = \max\left[ hd(\mathcal{S}_l, \mathcal{S}_{\tilde{l}}), hd(\mathcal{S}_{\tilde{l}}, \mathcal{S}_l) \right],
\end{equation}
where:
\begin{equation} \label{eq:hd}
    hd(\mathcal{S}_a, \mathcal{S}_b) = \max_{\mathbf{x}_i \in \mathcal{S}_a} \{\min_{\mathbf{x}_j \in \mathcal{S}_b}{\delta(\mathbf{x}_i, \mathbf{x}_j)}  \},
\end{equation}
and $\delta(\cdot,\cdot)$ denotes a Euclidean Distance between two points. From~\eqref{eq:hd} it can be observed that a change in the position of a single UE can have a significant impact on the Hausdorff Distance.

\subsection{Mean Distance}

A distance metric that can reduce the impact of a single UE on its outcome is the so-called Mean Distance. The idea is that first, a mean point is computed for each set of UEs' positions. Secondly, a Euclidean Distance is computed between these means:
\begin{equation} \label{eq:meabn_distance}
    d_{mean}(\mathcal{S}_l, \mathcal{S}_{\tilde{l}}) = \delta \left(\frac{\sum_{\mathbf{x}_i \in \mathcal{S}_l}\mathbf{x}_i}{|\mathcal{S}_l|}, \frac{\sum_{\mathbf{x}_j \in \mathcal{S}_{\tilde{l}}}\mathbf{x}_j}{|\mathcal{S}_{\tilde{l}}|} \right) 
\end{equation}
where $|\mathcal{S}_l|$, and  $|\mathcal{S}_{\tilde{l}}|$ stands for the number of elements in sets $\mathcal{S}_l$, and $\mathcal{S}_{\tilde{l}}$ respectively.

\subsection{Average Distance}

A similar approach is used in a metric called Average Distance. However, first, a Euclidean Distance is computed between every pair of points. Secondly, these distances are averaged:
\begin{equation} \label{eq:average_distance}
    d_{avg}(\mathcal{S}_l, \mathcal{S}_{\tilde{l}}) = \frac{1}{|\mathcal{S}_l||\mathcal{S}_{\tilde{l}}|} \cdot \sum_{\mathbf{x}_i \in \mathcal{S}_l, \mathbf{x}_j \in \mathcal{S}_{\tilde{l}}}{\delta(\mathbf{x}_i,\mathbf{x}_j)}
\end{equation}

\subsection{Sum of Minimums Distance} \label{subsec:sum_of_minimums_distance}

The slightly different idea lies under the Sum of Minimums Distance. First, for each UE position $\mathbf{x}_n$ from set $\mathcal{S}_l$ closest position from set $\mathcal{S}_{\tilde{l}}$ is found, and a related Euclidean Distance is computed. These distances are then summed. The same procedure is applied in opposite direction. The Sum of Minimums results in large distance values in relation to the remaining distance metrics considered in this paper, thus we decided to replace the sum with average. The final Sum of Minimums Formula is given by:
\begin{multline}\label{eq:sum_of_minimums_distance}
    d_{som}(\mathcal{S}_l, \mathcal{S}_{\tilde{l}}) = \frac{1}{2}\left( \frac{\sum_{\mathbf{x}_i \in \mathcal{S}_l}{\min_{\mathbf{x}_j \in \mathcal{S}_{\tilde{l}}}\delta(\mathbf{x}_i,\mathbf{x}_j)}}{|\mathcal{S}_l|}+ \right. \\ + \left. \frac{\sum_{\mathbf{x}_j \in \mathcal{S}_{\tilde{l}}}{\min_{\mathbf{x}_i \in \mathcal{S}_{l}}\delta(\mathbf{x}_j,\mathbf{x}_i)}}{|\mathcal{S}_{\tilde{l}}|}+ \right)
    \end{multline}
This distance metric can be thought of as a balance between the effect of extreme change in a single UE position, and the general structure of UEs' positions. It is similar to the Hausdorff Distance, yet in ~\eqref{eq:hausdorf_distance} $\max$ is replaced with average here. 
\section{Simulation Results} \label{sec:results}

For the purpose of evaluation of the distance metrics given in the previous section, a system-level simulator of MMIMO HetNet described in Sec.~\ref{subsec:network_architecture} is implemented. The simulator is described shortly below. A more detailed description can be found in~\cite{HOFFMANN2021}. 

\begin{figure}[htbp]
\centerline{\includegraphics[scale=0.22]{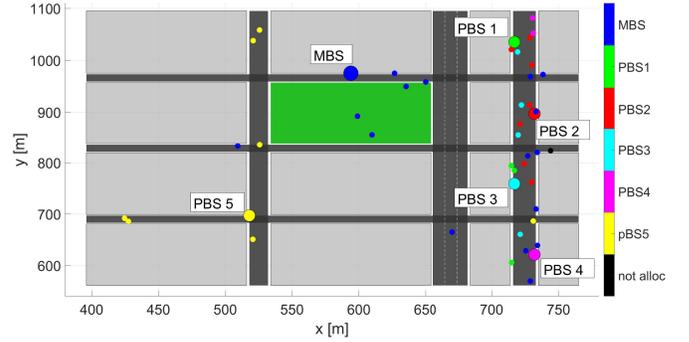}}
\caption{Deployment of the BSs (larger dots), together with one of the evaluated set of UEs positions (smaller dots).}
\label{fig:bs_deployment}
\end{figure}

The simulation area, deployment of BSs, and one of the evaluated sets of UEs positions are depicted in~Fig.~\ref{fig:bs_deployment}.  
There are 5 PBSs and a single MBS. They are equipped with full-digital antenna arrays, which is proper for advanced 5G and beyond networks~\cite{6gwp2020}. The antenna array installed at MBS consists of $128$ elements arranged in $16$ columns, and $8$ rows. Each of the PBSs is equipped with two antenna arrays of $16$ elements, having $2$ columns and $8$ rows. These antenna arrays are pointing in opposite directions. In each set of UEs positions, there would be $40$ UEs moving with the speed of $1.5$~m/s in random directions.

The considered MMIMO HetNet is OFDM-based, and exploits $15$ Modulation and Coding Schemes (MCSs). UEs are assigned to BSs with the use of the Dynamic Point Selection (DPS) performed every $10$~ms. This procedure can be finished within a single time slot~\cite{argwal2014dps}. We assume that each UE is served by the BS providing the highest RSS. If RSS observed by the UE is below the required threshold $P_{th}$ to every BS, it cannot be served by the network. For every UE connected to one of the active BSs, a procedure of scheduling and precoding is performed. The scheduler is designed to provide fairness, as described in~\cite{yoo2006}. The so-called Regularized-Zero-Forcing (RZF) precoder is used to enable the simultaneous allocation of the same time-frequency resources to various UEs. The BSs always allocates all radio resources to connected UEs, i.e., data stream follows the \emph{full-buffer} rule.

Radio channel coefficients are generated using 3D-Ray-Tracer, following the same procedure as in our previous papers. To provide significant changes in UEs' positions over the simulation period, the radio channel is generated in 15 batches lasting 60~ms, with 1~s long gaps between them. This procedure is introduced to reduce simulation time as explained in~\cite{HOFFMANN2021}.
The remaining simulation parameters, including e.g., power model, are summarized in Tab.~\ref{tab:simulation_parameters}.
\begin{table}[htbp]
\caption{Simulation Parameters}
\begin{center}
\begin{tabular}{|c|c|}
\hline
\textbf{Parameter} & \textbf{Value} \\
\hline
Simulation Time & ~15 s \\
\hline
Time Slot Duration & 0.5 ms \\ 
\hline
Number of UEs $N_{\mathrm{UE}}$ & 40 \\
\hline
 UEs Speed $v$ & $1.5 \frac{m}{s}$ \\
\hline
Number of PBSs $N_{\mathrm{BS}}-1$ & 5 \\ 
\hline
Central Frequency & $3.55$ GHz \\
\hline
Bandwidth & 300 MHz \\ 
\hline
Subcarrier Spacing & 30 kHz \\
\hline
Number of MBS Antennas & $128$ ($16\times 8$) \\
\hline
Number of PBS Antennas & $32$ ($2\times 2 \times 8$) \\
\hline
Precoder Type & RZF \\
\hline
RSS threshold $P_{\mathrm{th}}$ & -120 dBm \\
\hline
MBS Transmitted Power $P_{\mathrm{tx,MBS}}$ & 46 dBm \\
\hline
PBS Transmitted Power $P_{\mathrm{tx,PBS}}$ & 30 dBm \\
\hline
Amplifier Efficiency $\eta$ & 0.5 \cite{massivemimobook}\\
\hline
Transceiver Chain Power $P_{\mathrm{\widehat{TC}}}$ & 0.4 W \cite{massivemimobook}\\
\hline
Local Oscillator Power $P_{\mathrm{LO}}$ & 0.2 W \cite{massivemimobook} \\
\hline
BS Fix Power $P_{\mathrm{fix}}$ & 10 W \cite{massivemimobook} \\
\hline
BS Stand-By Power $P_{\mathrm{off}}$ & 10 W \cite{frenger2011} \\
\hline
\end{tabular}
\label{tab:simulation_parameters}
\end{center}
\end{table}

\subsection{Design of the Experiment}

This paper aims to compare four distance metrics, and their ability to match different sets of UEs' positions resulting in the same optimal RL action, i.e., set of active PBSs. To achieve this, we have generated a radio channel for 50 UEs. At first, a single subgroup of 40 UEs is chosen, and 30 simulation runs are performed repeating their motion pattern 
to learn, i.e., build REM with reliable knowledge. At this phase, UEs' positions are claimed to be accurately reported using RTK. For the REM-learning purpose, we used the ASR algorithm described in~\cite{HOFFMANN2021}. As a result of those 30 initial simulation runs, a REM of~15 entries is created. Next, 45 simulation runs were performed. Each simulation run was related to a new, random subgroup of 40 UEs out of 50 initially generated. After determination of the closest REM entry, using metrics defined in Sec.~\ref{sec:distance_metrics}, a greedy action was chosen~\eqref{eq:action_selection}, i.e., the configuration of PBSs which provided the highest EE in the past. The procedure was repeated for each distance metric. We ensured distance metrics are evaluated under identical unknown sets of UEs' positions by setting a constant random generator seed. Most importantly, this procedure was designed to test REM and its metrics, not the RL scheme. The experiment was conducted for two accuracy levels of positions reported by the UEs: RTK, having the error standard deviation $\sigma=1$~cm, and standard GPS receiver, with error standard deviation of  $\sigma=6$~m~\cite{hoffmann2020rtk}.

\subsection{Results Under RTK Localization Error}

\begin{figure}[htbp]
\centerline{\includegraphics[width=265pt]{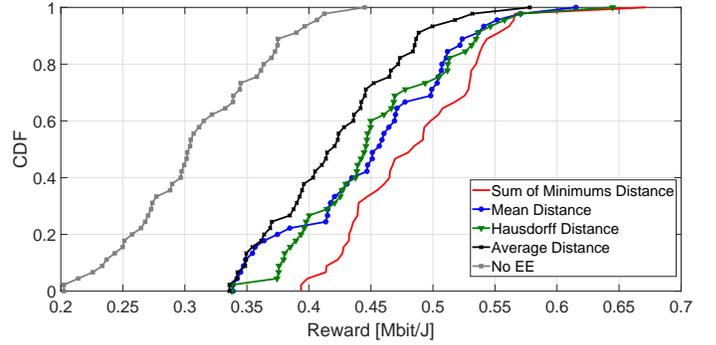}}
\caption{CDF of average EE achieved over randomly obtained sets of UEs positions, while using RTK for as a localization technique.}
\label{fig:cdf}
\end{figure}

The first results were obtained under RTK localization error, which in practice stands for almost perfect localization accuracy. The distribution of average EE achieved over 45 randomly obtained sets of UEs is depicted in Fig.~\ref{fig:cdf} in terms of Cumulative Distribution Function (CDF). 
At first, it can be observed that in all cases where historical knowledge from other sets of UEs positions stored in REM is utilized, network EE is improved over the No EE scenario where all PBs are active. The second major observation is that Sum of Minimums Distance~\eqref{eq:sum_of_minimums_distance} outperforms other distance metrics. Similarly, when REM utilizes the Average Distance~\eqref{eq:average_distance} it always provides the least EE gain. Between these two cases there are the Hausdorff Distance~\eqref{eq:hausdorf_distance} and Mean Distance~\eqref{eq:meabn_distance}. They seem to perform similarly. The same hierarchy of distance metrics is visible when considering results averaged over all tested sets of UEs positions depicted in Fig.~\ref{fig:bar_comparison}. A REM-based solution utilizing the Sum of Minimums Distance provides a $56\%$ improvement over a No EE scenario, while the one utilizing Average Distance gives only $36\%$.

The reason why the Sum of Minimums is the best, is because it balances sensitivity to a single UE movement and the overall structure of UEs' positions. On the other hand, Average Distance utilizes distances between every pair of UEs' positions and averages them. In this case, the result can be much affected by large distances between UEs being, e.g., on the opposite sides of the cell, yet close to other users.      

\subsection{Results Under GPS Localization Error}

\begin{figure}[htbp]
\centerline{\includegraphics[width=260pt]{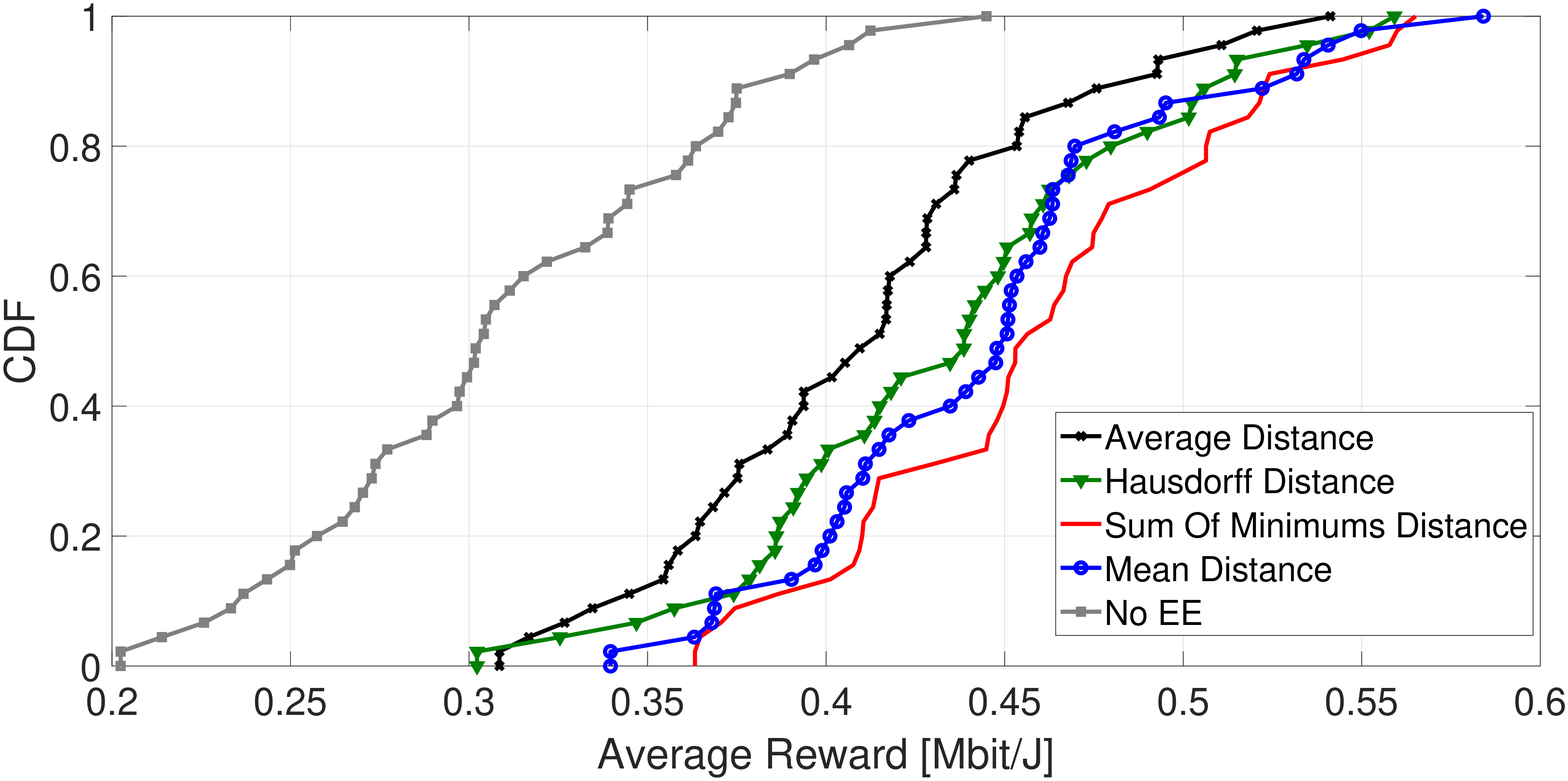}}
\caption{CDF of average EE achieved over randomly obtained sets of UEs positions, while using GPS as a localization technique.}
\label{fig:cdf_gps}
\end{figure}

\begin{figure}[htbp]
\centerline{\includegraphics[scale=0.26]{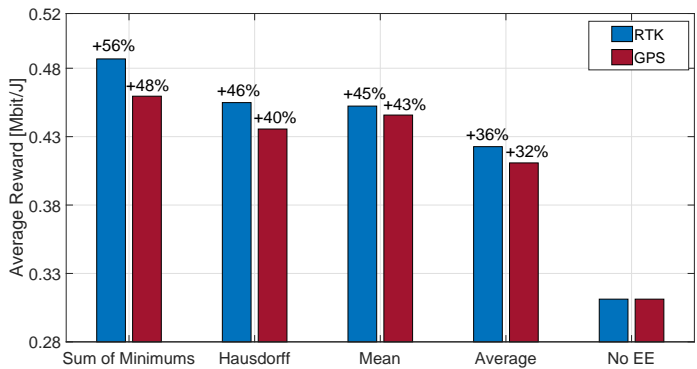}}
\caption{EE averaged over randomly obtained sets of UEs positions, in relation to the scenario without EE optimization--no EE.}
\label{fig:bar_comparison}
\end{figure}

A standalone GPS receiver is characterized by much poorer accuracy than RTK, i.e., standard deviation equals $6$~m instead of $1$~cm~\cite{hoffmann2020rtk}. In Fig.~\ref{fig:cdf_gps}, there is a CDF of achievable EE over 45 random sets of UEs positions, reported with Gaussian-distributed error introduced by GPS~\cite{misra2006global}. The relations between the performance of REM under different distance metrics are similar to the case utilizing RTK. However, now Mean Distance is better than Hausdorff Distance. It may be caused by the ability of Mean Distance to suppress localization error while computing the mean points of each set. At some points, it is even better than Sum of Minimums.
As it could be expected, the overall performance of REM is worse as a result of a less accurate localization method. However, differences are not very big. It can be seen in Fig.~\ref{fig:bar_comparison} that in the worst case of Average Distance, average EE gains over No EE scenario obtained using GPS are equal to $32\%$ instead of $36\%$ obtained using RTK. When considering the utilization of Sum of Minimums Distance average EE gains are reduced from $56\%$ to $48\%$. Nevertheless, results show that there is still a significant improvement in network EE while using historical knowledge from REM together with localization provided by GPS.   

\section{Conclusions} \label{sec:conclusions}

In this paper, we have compared 4 distance metrics in order to find out which one provides the best utilization of knowledge about the EE-maximizing set of active BSs stored in REM. The results have shown that the highest EE gains over a No EE scenario (all PBs are active) could be observed when the Sum of Minimums Distance was used. The EE is $56\%$, and $48\%$ higher than in the reference scenario while utilizing RTK and GPS as a localization technique, respectively. 

\bibliography{IEEEexample} 
\bibliographystyle{IEEEtran}

\end{document}